\documentclass[12pt]{article}
\pdfoutput=1
\usepackage{sbc-template}
\usepackage{graphicx,url}
\usepackage[utf8]{inputenc}
\usepackage{xcolor}

\usepackage{listings}
\usepackage[nolist]{acronym}
\usepackage{amsmath}
\usepackage[ruled,vlined,linesnumbered]{algorithm2e}
\usepackage{booktabs}
\usepackage{hyphenat}
\usepackage{tikz}
\usetikzlibrary{patterns}
\usepackage{pgfplots}
\pgfplotsset{compat=1.18}
\usetikzlibrary{external}

\title{Statistical Validation of Column Matching in the Database Schema Evolution of the Brazilian Public School Census}

\author{Muriki G. Yamanaka, Diogo H. de Almeida, Paulo R. \\Lisboa de Almeida, Simone Dominico, Leticia M. Peres,\\ Marcos S. Sunye, Eduardo C. de Almeida}
\address{Federal University of Paraná
  (UFPR) \\
   Centro de Computação Científica e Software Livre (C3SL)\\
   Curitiba -- PR --Brazil
  \email{\{dha21,mgy20,paulo,sdominico,lmperes,sunye,eduardo\}@inf.ufpr.br}
}

\begin{document}

\begin{acronym}[TDMA]
    \acro{CDF}{Cumulative Distribution Function}
    \acro{IQR}{Interquartile Range}
    \acro{LDE}{Educational Data Laboratory (\textit{Laboratório de Dados Educacionais})}
    \acro{MEC}{Ministry of Education}
\end{acronym}

\maketitle

\begin{abstract}
Publicly available datasets are subject to new versions, with each new version potentially reflecting changes to the data.
These changes may involve adding or removing attributes, changing data types, and modifying values or their semantics.
Integrating these datasets into a database poses a significant challenge: how to keep track of the evolving database schema while incorporating different versions of the data sources? 
This paper presents a statistical methodology to validate the integration of 12 years of open-access datasets from Brazil's School Census, with a new version of the datasets released annually by the Brazilian \ac{MEC}.
We employ various statistical tests to find matching attributes between datasets from a specific year and their potential equivalents in datasets from later years. The results show that by using the  Kolmogorov–Smirnov test we can successfully match columns from different dataset versions in about 90\% of cases.
\end{abstract}
 

\section{Introduction}

Integrating open data sources is a complex challenge in the development of web information systems.
Open data sources may exhibit structural changes over time when made public, including variations in data types, values, semantics, and missing values,  requiring constant evolution of the integrated database schema before the ingestion of new data~\cite{DBLP:books/daglib/0020812}.
The PRISM project reported an average of 217\% schema changes over a 48-month period across 12 large web information systems~\cite{DBLP:conf/hotswup/CurinoMZ09, DBLP:journals/vldb/CurinoMDZ13}.
For example, the Ensembl Genome project presented over 410 schema versions in 9 years. The Ensembl DB schema contains over 175 individual changes of primary and foreign keys in its schema evolution history.

The evolution of a database schema often leads to mapping errors, compromising the accuracy of stored data and ultimately leading to inconsistencies and inaccuracies in data analysis. 
Furthermore, differences in data presentation and evolving business needs can significantly hinder the incorporation of new data into existing databases.


In this paper, we introduce a statistical methodology to validate the integration of open-access datasets into the 
\ac{LDE} information system\footnote{
This work has received funding from the MEC/FNDE in the context of the Laboratório de Dados Educacionais (LDE) project (Grant agreement TED SIMEC No.: 11.437/2022).}. 
This methodology enables us to track the evolution of the system's database schema across different dataset releases. 
The \ac{LDE} system integrates open-access data from Brazil's School Census to support many studies and public educational policies~\cite{Schneider:2023, alves:2019, Schneider:2020, SILVEIRA:2021}. The \ac{LDE} database contains 12 years of School Census data and is freely accessible. Each year, \ac{MEC} publishes the School Census\footnote{Open data: https://www.gov.br/inep/pt-br/acesso-a-informacao/dados-abertos (in Portuguese)}, which includes comprehensive data from 179,500 schools, such as the number of students, teachers, and classes at each school. However, the publicly available data files have undergone 416 individual changes in naming conventions, as well as the addition and removal of columns over the years.  
These changes, driven by evolving government requirements, make it challenging for educational policymakers and researchers to access a unified and integrated reliable source.


Our methodology employs Goodness-of-fit statistical tests to evaluate the evolution of the LDE database schema enhancing the reliability of column matching. 
Goodness-of-fit tests are meant to define how well some sample of data fits with another given distribution \cite{DAgostino1986}. 
In our context of data integration, the tests conduct data profiling~\cite{DBLP:journals/vldb/AbedjanGN15}, analyzing column matching operations such as detecting additions, removals, and changes. This process helps minimize errors and inconsistencies in the evolution of an integrated database schema.



Overall, our main contribution in this paper are the following:

\noindent\textbf{Validation of the \ac{LDE} database schema:} Our methodology validates the quality of data integrated into the LDE database, thereby supporting the evolution of its schema.

\noindent\textbf{Quality metrics based on statistical tests:} Our methodology encompasses metrics from four goodness-of-fit statistical tests to evaluate the matches between the attributes of datasets from different releases: Kolmogorov-Smirnov test~\cite{bergerZhou2014}, Anderson-Darling test~\cite{10.1214/aoms/1177729437}, Welch's t-test~\cite{rayner2009}, and the F-test~\cite{hahsEtAl2020}.

\noindent\textbf{Analysis of the tests:}  we present the analysis of the results indicating that our methodology can correctly align the columns of different datasets in about the 90\% of cases considering the Top 3, and about 85\% considering the Top 1, showing high accuracy and effectiveness in the validation of the integrated schema.


This paper is structured as follows: Section~\ref{sec:lde} outlines the changes of the open-access data files and the potential integration problems in a database schema. Section~\ref{sec:meth} delineates the methodology used in this study. The findings are presented in Section~\ref{sec:results}, followed by a discussion of related work in Section~\ref{sec:rw}. Finally, Section~\ref{sec:conclusion} provides a summary of the study and outlines the next steps.


\section{Background}~\label{sec:lde}
Although schema evolution literature has long acknowledged the complexity of data source integration, the high computational costs associated with general schema evolution techniques have prevented their practical deployment~\cite{DBLP:conf/icse/CerqueusAS15, ScherzingerACAH16}.
Schema evolution refers to integrating changes to a data source over time, including adding new sources. Examples of source transformations over time include different column names, changes to the data domain, and their representation. It is also possible for columns or tables to be added as new sources are integrated~\cite{delplanque:2020}.



The \ac{LDE} system stores data from the School Census over the past 12 years, compiling a vast amount of educational information. Maintaining this dataset is crucial for monitoring trends over time and gaining valuable insights into the Brazilian educational context. Consequently, the \ac{LDE} system serves as a key resource for leveraging government open data in academic research.
Many projects maintained by \ac{MEC} and different universities depends on this data, such as the Cost-Student Quality Simulator (SIMCAQ)~\footnote{https://www.simcaq.c3sl.ufpr.br/ (in Portuguese)}~\cite{alves:2019}. SIMCAQ evaluates the cost of delivering quality education based on various educational and structural variables, such as class size, teacher salaries, and library resources. Another example of a system that depends on the LDE data is MapFor~\cite{Schneider:2023}, which tracks teachers' academic backgrounds. These projects demonstrably impact society, highlighting the importance of maintaining high data quality within the \ac{LDE} database.



There are many tools to assist the integration of datasets (a non-exhaustive survey on integration tools is found, here: \cite{DBLP:journals/vldb/CurinoMDZ13}). 
However, human intervention is often necessary to align open-access datasets containing historical information. 
This is further complicated by the evolution of the open data file structure over time, including changes in column names, value domains, and additions/removals of columns. This paper focuses on addressing the challenges associated with column name changes and additions/removals.

\begin{figure}[h]
    \centering
    \includegraphics[width=1\linewidth]{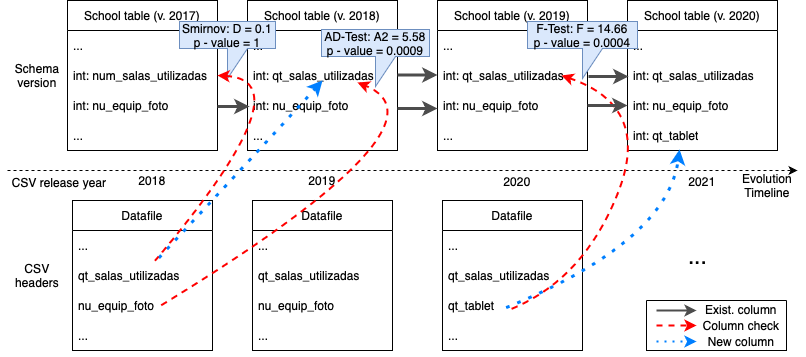}
    \caption{Illustration of schema evolution showing the data file headers from 2018 and 2020, as well as the impact of header changes on the integrated schema. Arrows indicate the mappings. Columns are presented in their original Portuguese names.} 
    \label{fig:schema_evolution}
\end{figure}

To illustrate the changes in column names, consider the following CSV headers released by the \ac{MEC} open-access files, shown in Figure~\ref{fig:schema_evolution}: ``$num\_salas\_utilizadas$'' and ``$qt\_salas\_utilizadas$''. 
To illustrate the introduction of new information, consider the column ``$qtde\_tablet$'' added in the 2020 data file. Finally, we illustrate columns when an attribute is no longer included in the data for a particular year. For example, the attribute ``$nu\_equip\_foto$'' was removed from the scholar census in 2017. 

Properly mapping all these changes in the \ac{LDE} database is essential to enhancing data quality. 
In Figure~\ref{fig:schema_evolution}, consider a scenario where new information is added to the data files of the scholar census, such as ``$qt\_salas\_utilizadas$''. Regardless of whether or not this information is already implied in the existing ``$num\_salas\_utilizadas$'' column, there might be a tendency to treat it as a new column. This could lead to the addition of a new column to the  \ac{LDE} database (``$qt\_salas\_utilizadas$''), resulting in schema evolution. However, mapping to this new column can make it difficult to infer existing information (``$num\_salas\_utilizadas$'') without detailed analysis. When a new column is created (such as ``$qt\_salas\_utilizadas$''), instances from previous years are filled with null values, and subsequent analysis may provide incorrect information, failing to indicate that previous data was present in another column (such as ``$num\_salas\_utilizadas$'').


\section{Goodness-of-fit Schema Evolution Methodology}\label{sec:meth}

In this section, 
we present our statistical methodology used in the integration of open access datasets into the \ac{LDE} database.
Our methodology employs goodness-of-fit statistical tests to match the CSV columns  released each year with the existing columns in the database. First, in Section \ref{sec:goodnessFit}, we define the goodness-of-fit tests in the context of schema evolution tests. In Section \ref{sec:proposed}, we define our matching algorithm, which uses specific metrics given by the goodness-of-fit tests to determine the correct match of each column for a given year.

\subsection{Goodness-of-Fit}\label{sec:goodnessFit}

Our main hypothesis is that Goodness-of-fit statistical tests ensure reliable \textit{data quality} metrics during column matching. These tests provide information about \textit{data distributions}, \textit{means}, \textit{variances}, and \textit{magnitude} of observed differences.
We use the well-known Kolmogorov–Smirnov test, Anderson–Darling test, Welch's t-test, and the F-test to compare the distributions of a column in a given year with possible matches from the next year.

Let $x:(x_1 ,x_2, \dots ,x_m )$ and $y:(y_1 ,y_2, \dots ,y_n)$ be the distributions (collected data between years) being compared of sizes $m$ and $n$, respectively. Also, let $\overline{x}$ and $\overline{y}$ be the means of $x$ and $y$, and $S_{x}^2$ and $S_{y}^2$ be the variances of $x$ and $y$. Now, we briefly describe each test.



\textbf{Kolmogorov–Smirnov test: } This test verifies if two samples are statistically similar. In our methodology, it determines whether the data in two columns from different years follow the same distribution. This allows us to evaluate data consistency over time by comparing the base year with the following year based on the distribution of the samples. Let $F_m$ and $G_n$ be the empirical \acp{CDF} for the $x$ and $y$ samples defined as follows:

\begin{equation}
    F_m(t) = \frac{\text{number of sample x\text{\textquotesingle}} \leq t}{m}
\end{equation}

\begin{equation}
    G_n(t) = \frac{\text{number of sample y\text{\textquotesingle}} \leq t}{n}
\end{equation}

\noindent the Kolmogorov-Smirnov test is defined as follows: 

\begin{equation}
D = max|F_m(t) - G_n (t)|, min(x, y) \leq t \leq max(x, y)
\end{equation}
\noindent where samples are considered to come from the same distribution if $D$ is small enough \cite{DAgostino1986,bergerZhou2014}. 
Considering the example illustrated by Figure~\ref{fig:schema_evolution}, the attributes ``$qt\_salas\_utilizadas$'' and ``$num\_salas\_utilizadas$'' present the same distribution and data type. In this particular case, the KS-test shows the $D = 0.1$ and $p-value=1$. 


\textbf{Anderson–Darling test: } Similar to the Kolmogorov-Smirnov test, the Anderson–Darling test considers the differences between the distributions, with the difference that the Anderson–Darling test gives more weight to the tails of the distributions when compared to the Kolmogorov-Smirnov test. For comparing two distributions, the Anderson–Darling statistic can be computed as follows:

\begin{equation}
    A^2 = \frac{1}{N(mn)}\sum_{j=1}^{N-1}\frac{(NX_j - jm)^2 + (NY_j - jn)^2}{j(N-j)}
\end{equation}

\noindent where $N = m + n$, $Z_1 < \dots < Z_N$ is the pooled ordered sample, and $X_j$ and $Y_j$ are the number of observations in $x$ and $y$ that are not greater than $Z_j$, respectively \cite{Pettitt:1976}. 
The Anderson–Darling test applied to  ``$nu\_equip\_foto$'' and ``$qt\_salas\_utilizadas$'' results in a statistic of 5.58 and a p-value of 0.0009, indicating a statistically significant difference in their distributions. This suggests that the columns are unlikely to be compatible.


\textbf{F-test: } 
By applying the F-test to each column, we compare the data variance across different data file versions (released in different years). If the test statistic $F$ is significant, it implies a substantial difference in data distribution, potentially indicating mismatches or missing data in the following year.
The F-test can be computed as follows: 

\begin{equation}
    F = \frac{S_{x}^2}{S_{y}^2}
\end{equation}

Values closer to 1 indicate similar variances, suggesting that both samples belong to the same distribution~\cite{hahsEtAl2020}. In our example, the F-test applied to ``$qtde\_tablet$'' and ``$qt\_salas\_utilizadas$'' columns yields a value of 14.66 and a p-value of 0.0004, indicating a statistically significant difference in their distributions.

\textbf{Welch's t-test}: The Welch's t-test is a variation of the Student's t-test, adapted to cases where the samples have unequal variances and sample sizes. The test is similar to the F-test, since it compares the variances. But unlike the F-test, this test considers the averaged values and sizes of the distributions. Its value is computed as follows:

\begin{equation}
    t = \frac{\overline{x} - \overline{y}}{\sqrt{\frac{S_{x}^2}{m} - \frac{S_{y}^2}{n}}}
\end{equation}

This variant of the t-test relaxes the assumption of equal variances between the two samples~\cite{rayner2009}. In our example, the attributes ``$qt\_salas\_utilizadas$'' and ``$num\_salas\_utilizadas$'' in Figure~\ref{fig:schema_evolution} demonstrate similar distributions, as evidenced by the Welch's t-test statistic of 0.069 and a high p-value of 0.95, indicating no significant difference in their distributions.

\subsection{The Schema Matching Algorithm}\label{sec:proposed}

Algorithm~\ref{alg:algColumnMatch} is designed to identify changes in data columns across different years. 
It compares columns from later years with the ``base schema'' (lines 2-5). The ``base schema'' represents the operational schema following a successful evolution and data integration. When the p-value from a comparison exceeds a specified threshold (line 7), the algorithm flags the column from the later year as a potential match and selects it as the best candidate for mapping to the corresponding base year column (line 8-10).

Most importantly, the algorithm enables the classification of data columns into three categories to guide integration decisions: identical columns, new columns, and columns without data.
Identical columns exhibit consistent data across different years.
New columns are introduced in the following year with no corresponding column in the base year.
Columns without data are present in the base schema but absent in the subsequent year.

The comparison of column matches falls under the broader domain of data profiling, which involves analyzing columns~\cite{DBLP:journals/vldb/AbedjanGN15, DBLP:journals/pvldb/PenaAN21}.
In data profiling, the number of potential column comparisons can grow exponentially with the number of attributes in a relation. While our algorithm inherits this complexity, it focuses on the specific task of comparing two columns, resulting in a worst-case scenario of quadratic complexity when dealing with identical schemas.

\begin{algorithm}[htpb]
	\footnotesize
	\SetKwInput{KwData}{Input}
	\KwData{$curr$: the current database columns; $new$: the new columns that arrived that must be matched; $gd$: goodness-of-fit method to be used; $p\_thresh$: minimum $p\_value$ to accept the column as a possible match.}
	\KwResult{A map matching the columns in $new$ with the columns in $curr$.}
    $matches = empty\ set$\\
	\For {$c\_column$ in $curr$}{
        $chosen\_col = NULL$\\
        $metric = NULL$\\
        \For {$n\_column$ in $new$}{
               $n\_metric, p\_value = gd(c\_column, n\_column)$\\
               \If {$p\_value \geq p\_thresh$}{
                   \If {($chosen\_col\ is\ NULL$ or $n\_metric$ is better than $metric$)}{
                        $metric = n\_metric$\\
                        $chosen\_col = n\_column$
                   }
               }
        }
        \tcp{the tuple (c\_column, n\_column) is a match between the current and the arrived column}
        $matches = matches \cup (c\_column, n\_column)$\\
        remove $chosen\_col$ from the set $new$
	}

    \tcp{Mark the remaining as no match (new columns)}
    \For {$n\_column$ in $new$}{
        $matches = matches \cup (NULL, n\_column)$
    }

    \Return matches
    
	\caption{\textsc{columnMatch}$(curr, new, gd, threshold)$.}
	\label{alg:algColumnMatch}
\end{algorithm}

\section{Experimental Results}~\label{sec:results}

In this section, we delve into the experiments we conducted using statistical techniques on data retrieved from the \ac{LDE} database. We describe the experimental protocol we followed and report the results obtained from the evaluation process.\footnote{ In this link, we provide access to data and the complete source code of the LDE system: https://dadoseducacionais.c3sl.ufpr.br/ (in Portuguese).}


\subsection{Experimental Protocol}

We used the R implementation for the Goodness-of-fit methods used in this work described in Section~\ref{sec:goodnessFit}. 
Some goodness-of-fit approaches may be sensible to outliers (e.g., the variance-based approaches, such as the F-Test), necessitating their removal during data preprocessing. We employed the \ac{IQR} method for outlier detection by finding the first (Q1) and third (Q3) quartiles, representing 25\% and 75\%, respectively. The IQR is the difference between Q1 and Q3. We identify and remove outliers as values falling below Q1 subtracted by $1.5$ times the IQR, or above Q3 added by $1.5$ times the \ac{IQR}. 

Instead of feeding the columns data directly to the Algorithm \ref{alg:algColumnMatch}, which could lead to problems when estimating the $p-value$ (indicating the confidence of a given column to be a correct match), we first transform the data of the columns in a \textit{10-bins} histogram. We defined $p\_thresh = 0.9$ (i.e., $\alpha = 0.1$) in Algorithm \ref{alg:algColumnMatch}, which is a common practice when accepting/rejecting the NULL hypothesis (i.e., the columns come from the same distribution).

\subsection{Results -- Matches Considering the Previous Year}\label{sec:resYearByYear}

In Table \ref{table:top1}, we show the results, considering the accuracy of the Top 1. We define a successful match as an exact match between a column and its corresponding column from the previous year. The accuracy is defined as $acc = \frac{hits}{\#columns}$. In Table \ref{table:top1}, column \textit{Year} defines the reference year, which we need to match the columns with the previous year. The column \textit{Changes} shows $[x]+$ as the number of new, $[y]-$ as the number of removed, and $[z]c$ as the number of changed columns when compared with the previous year (considering the ground-truth).

For example, consider 2019 in Table \ref{table:top1}. This year, when analyzing the official data made available from the \ac{MEC} and comparing it with the previous year (2018), no columns changed their names; 17 new columns appeared, presenting data that was not collected in the previous year, and eight columns disappeared, presenting data that was no longer collected.

\begin{table}[htpb]
    \centering
    \setlength{\tabcolsep}{5pt}
    \begin{tabular}{lcrrrr}
    \toprule
    Year & Changes & K–S Test & A-D Test & Welch's Test &  F-Test\\
    \midrule
     2007 & - & -   & -   & -   & -   \\
     2008 & \textit{no change}    & 0.667 & 0.333 & 0.5   &  0.333\\
     2009 & \textit{no change}    & 1.0   & 0.333   & 0.333 &  0.0\\
     2010 & \textit{no change}    & 0.833 & 0.667 & 0.333 & 0.333 \\
     2011 & \textit{no change}    & 0.333 & 0.667 & 0.667 & 0.167\\
     2012 & \textit{no change}    & 1.0   & 0.667 & 0.667 & 0.333\\
     2013 & $[0]c$ $[7]+$ $[0]-$  & 0.846 & 0.692 & 0.692 & 0.769 \\
     2014 & \textit{no change}    & 0.923 & 0.769 & 0.462 & 0.231 \\
     2015 & $[14]c$ $[0]+$ $[0]-$ & 1.0   & 0.846 & 0.538 & 0.308 \\
     2016 & \textit{no change}    & 1.0   & 0.846 & 0.385 & 0.462 \\
     2017 & \textit{no change}    & 0.615 & 0.538 & 0.462 & 0.231 \\
     2018 & $[7]c$ $[0]+$ $[0]-$  & 1.0   & 0.769 & 0.692 & 0.846 \\
     2019 & $[0]c$ $[17]+$ $[8]-$ & 0.824 & 0.882 & 0.706 & 0.706 \\
     2020 & \textit{no change}    & 1.0   & 1.0 & 0.636 &   0.591 \\
     2021 & \textit{no change}    & 0.818 & 0.727 & 0.364 & 0.273 \\ \hline
    \multicolumn{2}{l}{Average (stdev)} & 0.847 (0.189) & 0.701 (0.185) & 0.531 (0.139) &  0.393 (0.228)\\
    \bottomrule
    \end{tabular}
    \caption{Accuracy Considering the Top 1 results.}
    \label{table:top1}
\end{table}

As one can observe in Table \ref{table:top1}, the Kolmogorov–Smirnov (K-S) test presented the best results when considering the averaged results, correctly fitting the columns in over 80\% of the cases. The results in Table \ref{table:top1} also show that even in the event of many changes occurring in a given year, such as in the year 2019, where 19 new columns appeared and 2 columns disappeared, all tested approaches were able to correctly detect most of the correct fits, columns that should be created (new columns), and columns that should be disregarded (discontinued ones).

During the tests, we observed an interesting phenomenon: even when the column is not perfectly fitted with the previous year, the correct fit still presented a high probability (according to each test) of belonging to its proper fit. In light of this, we make a Top 3 analysis in Table \ref{table:top3}. The Top 3 consider a hit if the predicted fit appears in the three most probable fits for a given approach. The Top 3 can present a more realistic scenario than the Top 1 since it can show the most probable fits to a specialist, who will choose the correct one according to his/her domain knowledge.

\begin{table}[htpb]
    \centering
    \setlength{\tabcolsep}{5pt}
    \begin{tabular}{lcrrrr}
    \toprule
    Year & Changes & K–S Test & A-D Test & Welch's Test & F-Test \\
    \midrule
     2007 & - & -   & -   & -   & -   \\
     2008 & \textit{no change}    & 1.0    & 0.833 & 0.5   & 0.667 \\
     2009 & \textit{no change}    & 1.0    & 1.0   & 0.5   & 0.0   \\
     2010 & \textit{no change}    & 0.833  & 1.0 & 0.667 & 0.333   \\
     2011 & \textit{no change}    & 1.0    & 1.03 & 0.833 & 0.33   \\
     2012 & \textit{no change}    & 1.0    & 1.0   & 0.667 & 0.5   \\
     2013 & $[0]c$ $[7]+$ $[0]-$  & 0.846  & 0.692 & 0.462 & 0.538 \\
     2014 & \textit{no change}    & 0.846  & 0.846 & 0.846 & 0.462 \\
     2015 & $[14]c$ $[0]+$ $[0]-$ & 0.923  & 0.923 & 0.846 & 0.538 \\
     2016 & \textit{no change}    & 0.923  & 0.923 & 0.769 & 0.615 \\
     2017 & \textit{no change}    & 0.769  & 0.846 & 0.692 & 0.385 \\
     2018 & $[7]c$ $[0]+$ $[0]-$  & 1.0    & 1.0 & 0.846 & 0.923   \\
     2019 & $[0]c$ $[17]+$ $[8]-$ & 0.824  & 0.765 & 0.706 & 0.647 \\
     2020 & \textit{no change}    & 0.818  & 0.818 & 0.591 & 0.727 \\
     2021 & \textit{no change}    & 0.773  & 0.773 & 0.727 & 0.591 \\ \hline
    \multicolumn{2}{l}{Average (stdev)} & 0.897 (0.087) & 0.887 (0.101) & 0.689 (0.13) & 0.519 (0.21) \\
    \bottomrule
    \end{tabular}
    \caption{Accuracy Considering the Top 3 results.}
    \label{table:top3}
\end{table}

The Top 3 results in Table \ref{table:top3} show that both the Kolmogorov–Smirnov and Anderson–Darling tests correctly fit the columns in almost 90\% of the cases considering the Top 3 results. This result shows that these goodness-of-fit tests used in combination with our proposed Algorithm (Algorithm \ref{alg:algColumnMatch}) can significantly decrease the manual work of the domain's specialist, who will find the correct match in the first proposals of the algorithm instead of needing to find the proper fit considering all possible columns available for a given year.





\subsection{Results -- Matches Considering the Accumulated Years}\label{sec:resAccum}

In this Section, we follow the same protocol as in Section \ref{sec:resYearByYear}, with the difference that when trying to match the $nth$ reference year, all the data from the first year to the $nth - 1$ year are used to create the distribution to be compared. For instance, when trying to match the reference year of 2010, the data distribution from 2010 is compared with the distributions of the years 2007, 2008, and 2009 combined. To make this possible, we first normalize the histograms by dividing each bin by the number of tuples used to create the histogram.

The concept of this test is that by grouping more data to compare, we could get closer to the real underlying distribution of the data. The Top 1 results of this test are shown in Table \ref{table:top1 accumulated}.

\begin{table}[htpb]
    \centering
    \setlength{\tabcolsep}{5pt}
    \begin{tabular}{lcrrrr}
    \toprule
    Year & Changes & K–S Test & A-D Test & Welch's Test & F-Test \\
    \midrule
     2007 & - & -   & -   & -   & -   \\
     2008 & \textit{no change}    & 1.0   & 0.333 & 1.0   &  0.167\\
     2009 & \textit{no change}    & 0.667 & 0.667   & 0.333 &  0.0\\
     2010 & \textit{no change}    & 0.667 & 1.0   & 0.167 & 0.0   \\
     2011 & \textit{no change}    & 0.667 & 1.0   & 0.167 & 0.0   \\
     2012 & \textit{no change}    & 1.0   & 1.0   & 0.167 & 0.0   \\
     2013 & $[0]c$ $[7]+$ $[0]-$  & 1.0   & 1.0 & 0.615 & 0.462   \\
     2014 & \textit{no change}    & 0.923 & 0.692 & 0.231 & 0.077 \\
     2015 & $[14]c$ $[1]+$ $[0]-$ & 0.923 & 0.923 & 0.0   & 0.154 \\
     2016 & \textit{no change}    & 0.692 & 0.692 & 0.154 & 0.154 \\
     2017 & \textit{no change}    & 1.0   & 0.846 & 0.769 & 0.231 \\
     2018 & $[7]c$ $[0]+$ $[0]-$  & 1.0   & 0.846 & 0.846 & 0.769 \\
     2019 & $[0]c$ $[17]+$ $[8]-$ & 0.348 & 0.348 & 0.304 & 0.609 \\
     2020 & \textit{no change}    & 0.267 & 0.267 & 0.1   & 0.533 \\
     2021 & \textit{no change}    & 0.267 & 0.2   & 0.067 & 0.6   \\ \hline

    \multicolumn{2}{l}{Average (stdev)} & 0.744 (0.271) & 0.701 (0.287) & 0.351 (0.309) & 0.268 (0.26) \\
    \bottomrule
    \end{tabular}
    \caption{Accuracy Considering the Top 1 accumulated results.}
    \label{table:top1 accumulated}
\end{table}

As one can observe in Table \ref{table:top1 accumulated}, besides the good results of the Kolmogorov–Smirnov and Anderson–Darling tests, there was a decrease in the accuracy for all the tested approaches when compared with the results when the columns were matched considering only the previous year (Section \ref{sec:resYearByYear})\footnote{We identified a similar behavior in the Top 3 results, which are not included in this work due to length restrictions.}. We hypothesize that external factors, such as data collection policy changes over time, made it more difficult to make a correct match when including the data from previous years in the distribution for comparison. The Kolmogorov-Smirnov test corroborates this hypothesis since the results got worse for the latest years when compared with Section \ref{sec:resYearByYear}, since in the latest years, there were more old data aggregated in the columns to make the comparison.

\section{Related Work}\label{sec:rw}


Schema evolution management has been the focus of several works over the years. These works conduct empirical investigations into relational schema evolution~\cite{Qiu:2013, Vassiliadis:2015, Vassiliadis:2017, Vassiliadis:2019}. In~\cite{Klettke:2017}, the authors evaluate schema evolution histories over time, examining data from a data lake and schema versioning. Our work analyzes data integration quality and tracks the evolution of the database schema.

Some works have evaluated the schema evolution in the NoSQL database~\cite{Meurice:2017, Ringlstetter:2016}. In~\cite{DBLP:conf/icse/CerqueusAS15},  the authors discuss the implementation and customization of verification rules to help developers manage schema evolution and prevent compatibility issues and data loss. ~\cite{Scherzinger:2020, ScherzingerACAH16, CerqueusA15} investigate the evolution of NoSQL database schema, focusing on their flexibility, denormalization practices, and changes over development time through empirical analysis of open-source projects. Our work evaluates the quality of schema evolution in a relational database, which implies distinct challenges. The NoSQL schema evolution has greater flexibility and denormalization. However, relational databases enforce constraints and a greater need to maintain integrity.

Prism/Prism++~\cite{DBLP:conf/hotswup/CurinoMZ09, DBLP:journals/vldb/CurinoMDZ13} implements a solution focused on schema evolution in relational databases. Prism uses the data dictionary to track changes in the data schema. It describes an integrated solution to predict and evaluate the impact of schema changes and integrity constraints. The objective is to minimize downtime by automating database migration and documenting schema evolution.

The work of~\cite{delplanque:2020} discusses the challenges of evolving relational database schema. The authors propose a meta-model approach to automate modifications after database changes, providing recommendations to maintain a consistent state.
As observed in~\cite{etien2024}, a meta-model for analyzing the impact of changes and ensuring database relational constraints are verified. In contrast, our methodology evaluates the data distribution and other statistical measures without analyzing attribute names. 
This approach allows us to monitor schema evolution from a data-centric perspective, providing an understanding of how data changes over time.

\section{Conclusion and Future Work}\label{sec:conclusion}

In this paper, we presented a methodology for identifying matches between attributes of the census datasets from a given year and their possible matches in a dataset released in a subsequent year. Our hypothesis was that using statistical tests to evaluate the evolution of the \ac{LDE} database schema enhances the reliability of column matching. Indeed, our methodology finds the matching attributes and shows what the changes are, such as adding new data or data columns not present in the year evaluated. Results showed that our approach, combined with the Kolmogorov–Smirnov test, significantly reduced the manual effort required by specialists. In our methodology, specialists can identify the correct matches rather than having to evaluate all possible columns available for a given year.

As future work, we plan to evaluate attributes that have binary and categorical data. These data require the application of different statistical tests than those used for numerical data. Besides, we also plan to automate the analysis of schema quality as necessary by evaluating column matching. At this point, we intend to study the use of machine learning to facilitate the integration of databases and manage schema evolution.


\bibliographystyle{sbc}
\bibliography{sbc-template}
\end{document}